\documentclass[a4paper,11pt]{article}
\usepackage{pos}
\usepackage{hyperref,setspace,subcaption}
\usepackage{mathtools}

\DeclareMathOperator{\sgn}{sgn}   
\DeclarePairedDelimiter\abs{\lvert}{\rvert}%

\title{Data-driven analysis for understanding ultrahigh energy cosmic ray source spectra}
 \ShortTitle{Data-driven analysis for understanding UHECR source spectra}

\author*[a,b,c]{Marco Stein Muzio}
\author[d,e,f]{Luis A. Anchordoqui}
\author[g]{Michael Unger}

\affiliation[a]{Department of Astronomy and Astrophysics, Pennsylvania
  State University, University Park, PA 16802, USA} 
\affiliation[b]{Department of Physics, Pennsylvania State University, University Park, PA 16802, USA}
\affiliation[c]{Institute of Gravitation and the Cosmos, Center for Multi-Messenger Astrophysics, Pennsylvania State University, University Park, PA
16802, USA}

\affiliation[d]{Department of Physics and Astronomy,  Lehman College, City University of
  New York, NY 10468, USA
}
\affiliation[e]{Department of Physics,
 Graduate Center, City University
  of New York,  NY 10016, USA
}
\affiliation[f]{Department of Astrophysics,
 American Museum of Natural History, NY
 10024, USA
}

\affiliation[g]{Institut f\"ur Astroteilchenphysik, Karlsruher Institut f\"ur Technologie, Karlsruhe 76344, Germany}

\emailAdd{msm6428@psu.edu}

\abstract{One of the most challenging open questions regarding the origin of ultrahigh energy cosmic rays (UHECRs) deals with the shape of the source emission spectra. A commonly-used simplifying assumption is that the source spectra of the highest energy cosmic rays trace a Peters cycle, in which the maximum cosmic-ray energy scales linearly with $Z$, i.e., with the charge of the UHECR in units of the proton charge. However, this would only be a natural assumption for models in which UHECRs escape the acceleration region without suffering significant energy losses. In most cases, however, UHECRs interact in the acceleration region and/or in the source environment changing the shape of the source emission spectra. Energy losses are typically parameterized in terms of $Z$ and the UHECR baryon number $A$, and therefore one would expect the source emission spectra to be a function of both $Z$ and $A$. Taking a pragmatic approach, we investigate whether existing data favor any region of the $(Z,A)$ parameter space. Using data from the Pierre Auger Observatory, we carry out a maximum likelihood analysis of the observed spectrum and nuclear composition to shape the source emission spectra for the various particle species. We also study the impact of possible systematic uncertainties driven by hadronic models describing interactions in the atmosphere.}

\ConferenceLogo{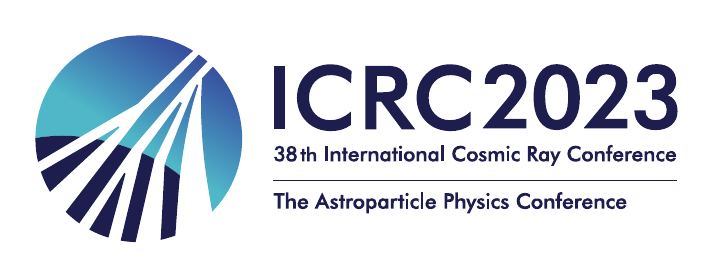}

\FullConference{%
38th International Cosmic Ray Conference (ICRC2023)\\
  26 July - 3 August, 2023\\
  Nagoya, Japan}

\begin{document}
\maketitle

\section{Introduction}

\par
Among the many open questions in the study of ultrahigh energy cosmic rays (UHECRs, $E \gtrsim10^{18}~\mathrm{eV} = 1~\mathrm{EeV}$) is the dependence of the maximum energy of nuclei produced by sources on their mass $A$ and charge $Z$. To simplify modeling of the UHECR spectrum and composition, one often assumes nuclei are accelerated to a common maximum rigidity $R_\mathrm{max}$, i.e.\ nuclei follow a \textit{Peters cycle}~\cite{PetersCycle}, so that $E_\mathrm{max}^A \propto Z$. 

\par
However, other possible scalings of the maximum energy with $(A,Z)$ exist depending on the details of the acceleration mechanism and the dominant energy loss processes to which UHECRs are subject. For example, synchrotron and curvature radiation loss rates scale as $Z^4/A^2$ and $Z^2$, respectively~\cite{Ptitsyna:2008zs,Rieger:2011ch,Anchordoqui:2018qom}. When UHECRs are diffusively accelerated, significant synchrotron losses lead to a maximum energy which scales as $A^4/Z^4$~\cite{Rieger:2011ch}. On the other hand, when UHECRs undergo a one-shot acceleration process, synchrotron losses lead to a $A^2/Z^{3/2}$ scaling, whereas curvature radiation losses produce a $A/Z^{1/4}$ scaling~\cite{Ptitsyna:2008zs}. Photodisintegration processes, which have been explored extensively in~\cite{Unger:2015laa,Muzio:2019leu,Muzio:2021zud}, preserve the energy-per-nucleon of the primary CR, so that $E^A_\mathrm{max} \propto A$. Finally, beyond the Standard Model scenarios may result in a universal maximum energy scale~\cite{Montero:2022prj,Anchordoqui:2022ejw,Noble:2023mfw}, which would predict that the maximum energy of nuclei is independent of their mass or charge.

\par
We aim to explore the degree to which the UHECR data favors or disfavors these alternative scenarios to the traditional Peters cycle assumption. We also discuss the observational signatures of alternative scenarios which might be used to distinguish them from a Peters cycle.

\section{Model}

\par
To study the degree to which current UHECR data can distinguish different scenarios for the dependence of the maximum energy on $(A,Z)$ we adopt a simple two-parameter model:
\begin{align}~\label{eq:Emax}
    E_\mathrm{max}^A = E_0 Z^\alpha A^\beta~,
\end{align}
\noindent
where $E_0$ corresponds to the maximum proton energy. Within this model a Peters cycle would be given by $(\alpha,\beta)= (1,0)$. 

\par
We use \eqref{eq:Emax} to set the maximum energy scale of nuclei escaping a standard UHECR source, which we assume to follow a star-formation rate (SFR) evolution~\cite{Robertson:2015uda}. In particular, we adopt a simplified model with five mass groups escaping the source, representing $p$, He, CNO, Si, and Fe, each following a exponentially-cutoff single power-law spectrum: $J_A\propto E^\gamma \exp(-E/E_\mathrm{max}^A)$. The relative abundances of these mass groups are free parameters in the fit. This leaves only two free parameters: the maximum proton energy, $E_0$, and the spectral index of the escaping spectra, $\gamma$, which we take to be common among all the mass groups. For a given set of parameters, UHECRs are propagated to Earth accounting for their interactions with the cosmic microwave background (CMB) and extragalactic background light (EBL) using propagation matrices built from CRPropa3~\cite{AlvesBatista:2016vpy}.

\par
The predicted spectrum and composition at Earth are fit to data from the Pierre Auger Observatory (Auger)~\cite{Yushkov:2020nhr,PierreAuger:2021hun}. In particular, we consider two hadronic interaction models, \textsc{Sibyll2.3d}~\cite{Riehn:2019jet} and \textsc{EPOS-LHC}~\cite{Pierog:2013ria}, to interpret the depth of shower maximum, $X_\mathrm{max}$, data in terms of $\ln{A}$. Our model parameters are then tuned to minimize the total $\chi^2$:
\begin{align}
    \chi^2 = \displaystyle \sum_i \frac{(J_i-J_{m,i})^2}{\sigma_{J,i}^2} + \sum_i \frac{(\mu_i-\mu_{m,i})^2}{\sigma_{\mu,i}^2} + \sum_i \frac{(V_i-V_{m,i})^2}{\sigma_{V,i}^2} ~,
\end{align}
\noindent
where $J$ is the UHECR flux, $\mu$ and $V$ are the mean and variance of $\ln{A}$ respectively, and the subscript $m$ denotes the model prediction. We perform this fit to the Auger data above $10^{18.8}$~eV, which given our free parameters leaves $N_\mathrm{dof}=29$. In a coming publication we also consider the sensitivity of our results to systematic shifts of the data, but here we focus on our benchmark set of data shifts, which provide the best-fit to the Auger data overall: shifting the energy scale by $\mathrm{dlg}E=+0.1$ and shifting the $\langle X_\mathrm{max} \rangle$ by $-1\sigma_X$.

\begin{figure}
    \centering
    \begin{subfigure}{0.7\linewidth}
        \includegraphics[width=\textwidth]{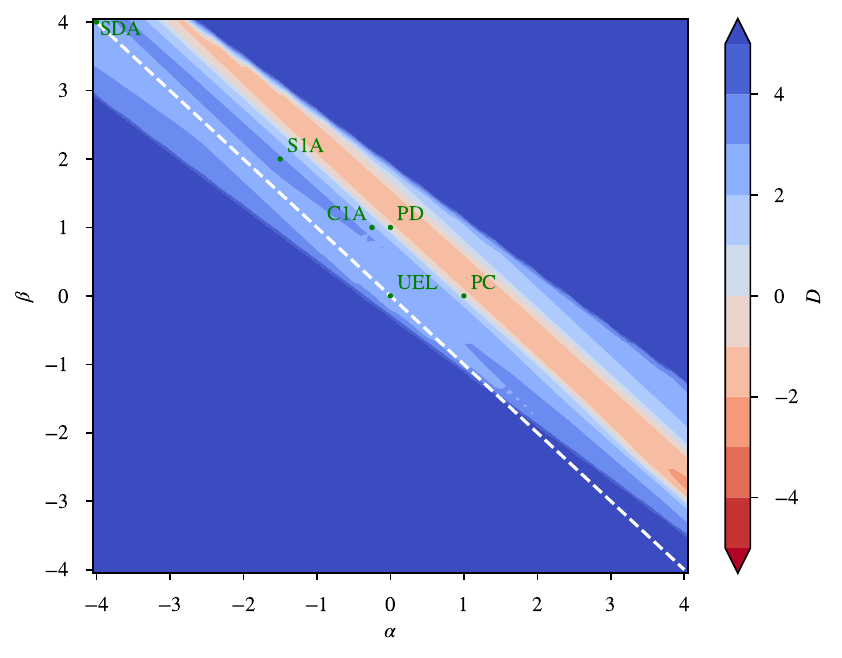}
        \caption{}
        \label{fig:benchmark_sibyll}
    \end{subfigure}
    \begin{subfigure}{0.7\linewidth}
        \includegraphics[width=\textwidth]{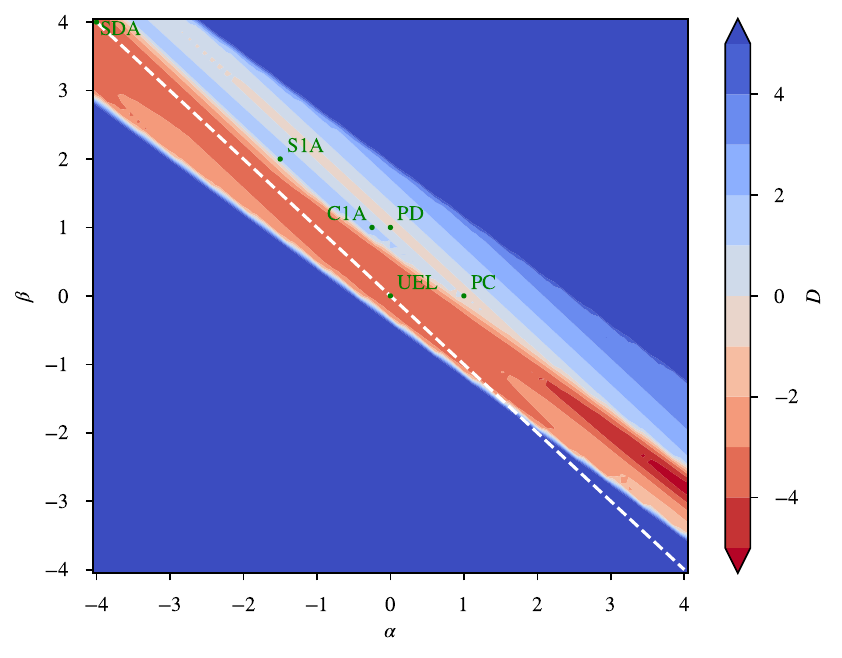}
        \caption{}
        \label{fig:benchmark_epos}
    \end{subfigure}
    \caption{Change in quality of fit to the UHECR spectrum and composition relative to a Peters cycle. We consider (a) \textsc{Sibyll2.3d} and (b) \textsc{EPOS-LHC}. The family of scenarios with $\alpha + \beta = 0$ are indicated by the white dashed line. The Peters cycle and a number of alternative scenarios are highlighted (green dots).}
    \label{fig:benchmark}
\end{figure}

\section{Results}

\par
To assess the degree to which an alternative scenario, $(\alpha,\beta)$, is favored or disfavored with respect to a Peters cycle, $(1,0)$, we use
\begin{align}
    D \equiv \sgn\left(\chi_{\alpha,\beta}^2 - \chi^2_\mathrm{Peters}\right) S^{-1} \sqrt{\abs*{\chi_{\alpha,\beta}^2 - \chi^2_\mathrm{Peters}}}
\end{align}
\noindent
as a metric, where $S = \sqrt{\min(\chi_{\alpha,\beta}^2,\chi^2_\mathrm{Peters})/N_\mathrm{dof}}$. With this definition, $D>0$ indicates a worse fit than a Peters cycle and $D<0$ indicates the fit has improved compared to a Peters cycle. The statistical significance of the change in fit quality relative to a Peters cycle can be calculated from $D$ using Wilks' theorem. Figure~\ref{fig:benchmark} shows $D$ in the $\alpha-\beta$ plane and highlights the Peters cycle (PC) and a number of alternative scenarios: a photodisintegration-limited spectrum (PD), a synchrotron-limited diffusion accelerated spectrum (SDA), a synchrotron-limited one-shot accelerated spectrum (S1A), a curvature radiation-limited one-shot accelerated spectrum (C1A), and a universal energy loss spectrum (UEL). 

\par
Different values of $(\alpha,\beta)$ change the relative energies of the different mass groups escaping the source, which in turn changes the quality of fit. We expect that directions in the $\alpha-\beta$ plane along which the ratio of maximum energies of different nuclei is constant will produce similar quality fits to the UHECR data. In practice, this is realized due to the degeneracy between $A$ and $Z$: for stable nuclei $A\simeq 2Z$, while for protons $A=Z=1$. Combining this fact with \eqref{eq:Emax} allows one to write the ratio of maximum energies between nuclei as
\begin{align}\label{eq:nucleiRatio}
    \frac{E_\mathrm{max}^A}{E_\mathrm{max}^{A'}} = \left( \frac{A}{A'}\right)^{\alpha+\beta} ~,
\end{align}
\noindent
and between a nucleus and a proton as
\begin{align}\label{eq:ApRatio}
    \frac{E_\mathrm{max}^A}{E_\mathrm{max}^p} = 2^{-\alpha} A^{\alpha+\beta}~.
\end{align}
\noindent
Therefore, expect directions where $\alpha+\beta$ is constant to produce similar quality fits. However, if a substantial proton component exists in the escaping spectrum then directions along which $(1-\log_{A}{2})\alpha + \beta$ is constant will produce similar quality fits. This expectation can be seen clearly in Fig.~\ref{fig:benchmark}.

\par
\begin{figure}[htpb!]
    \centering
    \includegraphics[width=0.7\linewidth]{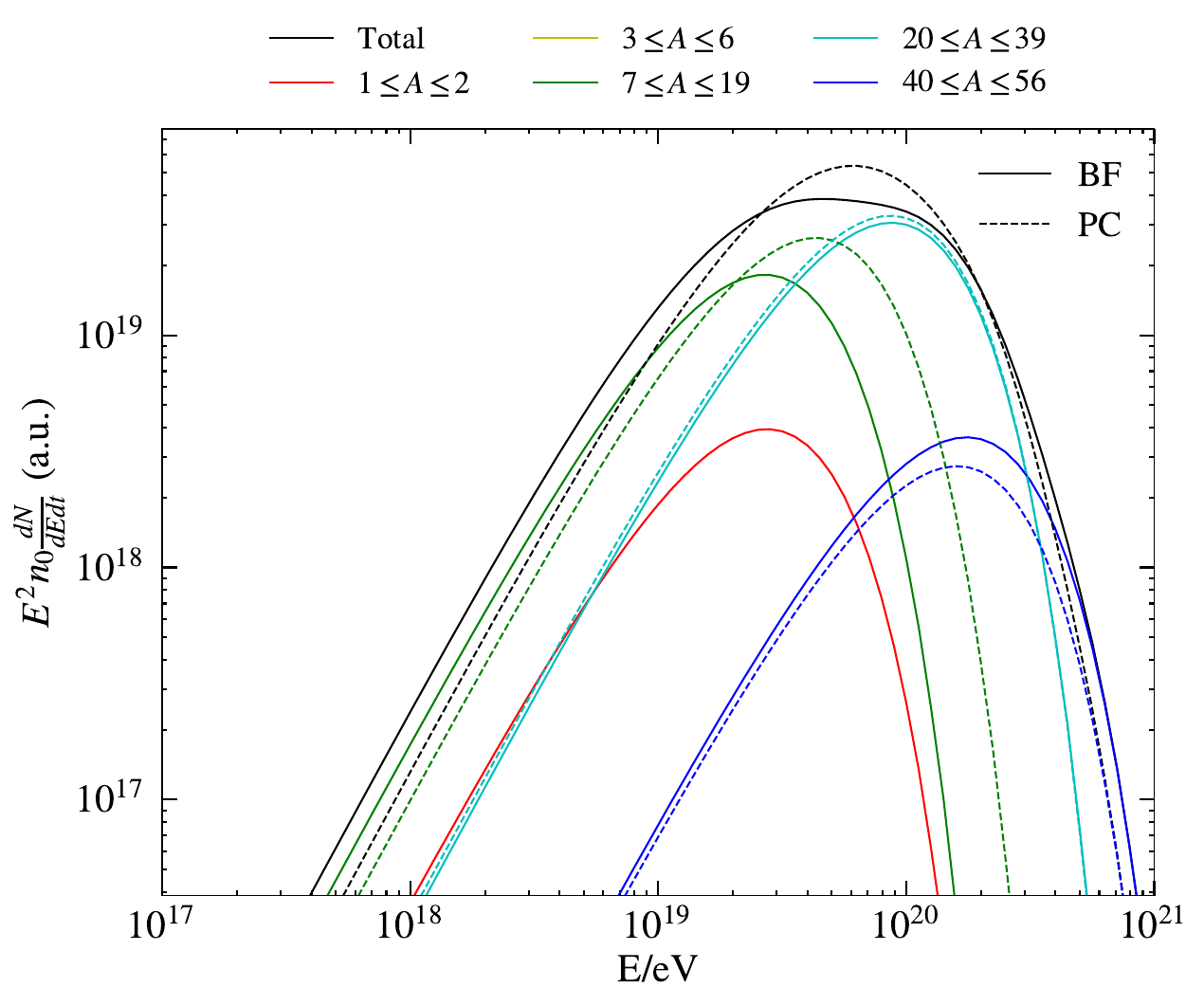}
    \caption{Best-fit escaping spectra for the best-fit ($\alpha\simeq 6.75$, $\beta\simeq -5$) scenario (solid lines) compared to those for a Peters cycle (dashed lines) under \textsc{EPOS-LHC}. Spectra are broken down by mass group (colored lines).}
    \label{fig:bestfit_esc}
\end{figure}

Equations~\eqref{eq:nucleiRatio} and~\eqref{eq:ApRatio} have two important consequences. First, equation~\eqref{eq:nucleiRatio} implies that if no significant proton component exists in the escaping spectrum, only the value of $\alpha+\beta$ impacts the fit. Therefore, in this circumstance different acceleration scenarios fall into families which share a common value of $\alpha+\beta$. Second, in the circumstance that a significant proton component is present in the escaping spectrum, equation~\eqref{eq:ApRatio} indicates that this proton component can peak either below or above the peak energy of nuclei, depending on the value of $\alpha$. In particular, this means that while for $\alpha+\beta>0$ nuclei will have peak energies which are order according to their mass (i.e. He peaking at the lowest energies and Fe at the highest), the proton component need not obey this ordering. Large values of $\alpha$ will, therefore, place the peak of the proton component towards the iron-end of the UHECR spectrum (as is illustrated in Fig.~\ref{fig:bestfit_esc}).

\par
Regardless of the hadronic interaction model considered, Fig.~\ref{fig:benchmark} shows alternative scenarios from the Peters cycle can produce a better fit to the data. This result holds for all alternative data shifts explored as well. In some cases the improvement in fit can result in values of $D<-5$. 

\par
The region producing the largest improvement over a Peters cycle in the case of \textsc{EPOS-LHC} is one where $(1-\log_{A}{2})\alpha + \beta$ is constant, rather than $\alpha+\beta$ (as evidenced by this region not being parallel to the white dashed line representing $\alpha+\beta=0$). This implies that there is a significant proton component escaping the source in the scenarios producing the best-fits to the UHECR spectrum and composition. In particular, we find this region roughly follows $\beta \simeq 0.4 - 0.8\alpha$, indicating that it is the relation of the maximum proton energy to the maximum energy of $A\simeq 32$ which is driving the fit. 

\par
\begin{figure}[htpb!]
    \centering
    \includegraphics[width=0.7\linewidth]{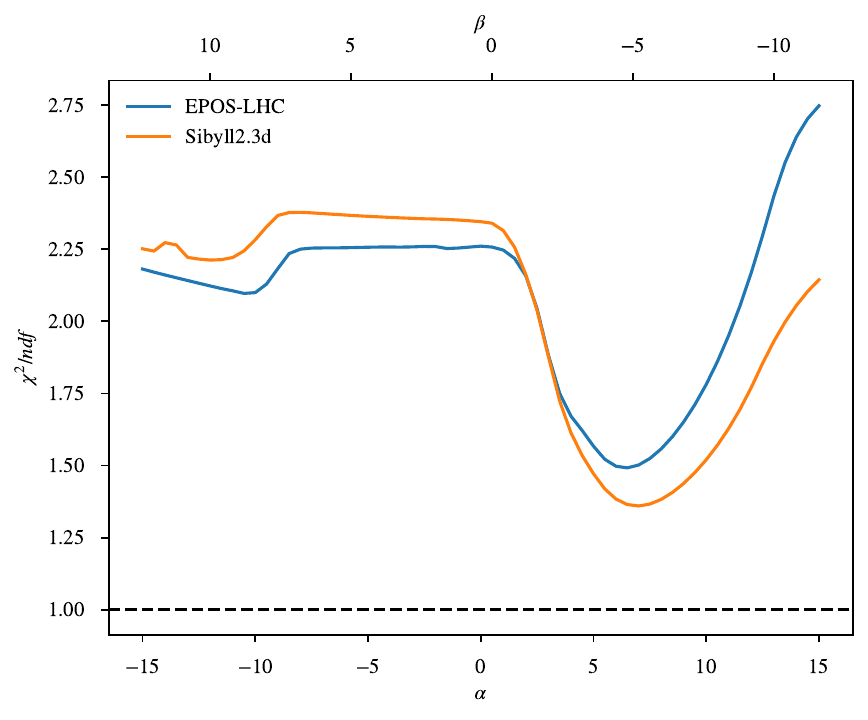}
    \caption{Reduced $\chi^2$ along the line $\beta = 0.4 - 0.8 \alpha$ for \textsc{Sibyll2.3d} and \textsc{EPOS-LHC}. In both cases the minimum appears around $\alpha\simeq 6.5$ and $\beta \simeq = -5$.}
    \label{fig:1Dscan}
\end{figure}

It is clear from Fig.~\ref{fig:benchmark} that the best-fit lies outside of the plotted range (which was driven by the range of $\alpha$ and $\beta$ values among the reference set of alternative scenarios). To explore how far the best-fit is outside this range we performed a 1-D scan along the line following the best-fit region for \textsc{EPOS-LHC}: $\beta \simeq 0.4 - 0.8\alpha$. The results of this scan are shown in Fig.~\ref{fig:1Dscan}. The best-fit occurs at $\alpha\simeq 6.75$ and $\beta\simeq -5$ for both hadronic interaction models considered. Currently, we are unaware of any processes, observed or theoretical, which could produce such a scaling for the UHECR maximum energy. For the reader's reference, in Fig.~\ref{fig:bestfit_esc} we plot the escaping spectra for this best-fit (BF) model compared to those for a Peters cycle, assuming \textsc{EPOS-LHC}.

\par
While it is difficult to distinguish between the possibilities explored above using current UHECR data, there are some distinct observational signatures of alternative scenarios to a Peters cycle which could be used in the future. The most straight forward way to determine which scenario most observed UHECRs fall into would be to measure the peak energies of each mass group. This would directly probe $(\alpha,\beta)$, but is difficult to measure in practice. Here we consider two alternative signatures of alternative scenarios.

\par
First let us consider the case that there is a substantial proton flux escaping UHECR sources. In this case, if the value of $\alpha\neq1$ then this proton component will peak at energy a factor of $2^{\alpha-1}$ different than the Peters cycle expectation. In particular for large values of $\alpha$ this component will peak at energies higher than He. Therefore, measurement of the peak energy of the proton component of the UHECR spectrum will constrain the value of $\alpha$. Moreover, if the value of $\alpha$ is large enough, so protons escaping the source will exceed the GZK threshold~\cite{Greisen:1966jv,Zatsepin:1966jv} and will therefore produce a substantial flux of EeV cosmogenic neutrinos. Measurement of such neutrinos will also provide a probe of $\alpha$. 

\par
In the case where no substantial flux of protons escapes the source only families of scenarios (those with a common value of $\alpha + \beta$) can be distinguished from one another. In particular, the Peters cycle family of scenarios ($\alpha+\beta=1$) can be distinguished from alternative families by measuring the proton component of the UHECR spectrum. These protons, by assumption, are not produced by the source directly and instead are the product of heavier nuclei photodisintegrating off of the CMB and EBL. These interactions preserve the energy-per-nucleon of the primary CR so that protons produced through photodisintegration of a CR with mass $A$ will have a peak energy of 
\begin{align}
    E_{A\mathrm{PD, max}}^p = 2^{-\alpha} E_0 A^{\alpha+\beta-1}~.
\end{align}
\noindent
For the Peters cycle family of scenarios, all photodisintegrated protons will have the same peak energy irrespective of their parent CR's mass. However, for alternative scenarios this energy will depend on the mass of the primary CR, so that the ratio of their maximum energies is given by
\begin{align}\label{eq:PDratio}
    \frac{E^p_{A\mathrm{PD,max}}}{E^p_{A'\mathrm{PD,max}}} = \left( \frac{A}{A'} \right)^{\alpha+\beta-1}~.
\end{align}
\par
Equation~\eqref{eq:PDratio} implies that the spectrum of protons may have multiple peaks and will be much more extended in energy than would be expected from a Peters cycle. Therefore, measurement of an extended proton component will be a smoking gun signature of an alternative scenario to a Peters cycle.

\par
An important caveat to keep in mind is that fundamentally all of the signatures discussed rely on the measurement of an unexpectedly energetic proton component or their secondary neutrinos. This can be mimicked by a second population of UHECRs which produces a large flux of protons at higher energies than the population producing the bulk of observed UHECRs. The implications of this possibility have been explored in a number of studies including~\cite{vanVliet:2019nse,Muzio:2023skc,Ehlert:2023btz}.

\section{Summary}

\par
While a Peters cycle has been a convenient simplifying assumption in the study of UHECRs, there are a number of alternative scenarios motivated by both well-known and beyond the Standard Model processes. The UHECR data today cannot firmly establish which of these scenarios is realized by Nature, but some alternative scenarios are able to describe current UHECR spectrum and composition data better than is possible using the classic Peters cycle assumption. In particular, exotic scenarios can improve fits to UHECR data at a high level of significance. There are a number of observational signatures for alternative scenarios, including a proton component extending across a large energy range and the production of cosmogenic neutrinos, which can be used to constrain these possibilities. Until then, we must keep in mind the possibility that Nature may provide a UHECR spectrum that is more rich than a simple Peters cycle.  

\clearpage

\acknowledgments

The work of L.A.A. is supported by the U.S. National Science
Foundation (NSF Grant PHY-2112527). The research of M.S.M. is supported by the NSF MPS-Ascend Postdoctoral Award \#2138121. 

\begingroup
\bibliographystyle{ICRC}
\bibliography{references}
\endgroup

\end{document}